\documentclass[aps,prl,twocolumn,showpacs,superscriptaddress,groupedaddress,floatfix,preprintnumbers]{revtex4}
\usepackage{graphicx}

\usepackage{subfig}
\usepackage{color}
\usepackage{chemmacros}
\usepackage{filecontents}
\usepackage[titletoc,toc,title]{appendix}
\usepackage[font=small,labelfont=bf,
   justification=justified,
   format=plain]{caption}
   
   \newcommand{\Kelvin}{\mathrm{K}}
\newcommand{\perKelvin}{\mathrm{K^{-1}}}
\newcommand{\mol}{\mathrm{mol}}
\newcommand{\permol}{\mathrm{mol^{-1}}}
\newcommand{\perdmcube}{\mathrm{dm^{-3}}}
\newcommand{\perm}{\mathrm{m^{-1}}}

\newcommand{\pernmsquare}{\mathrm{nm^{-2}}}

\newcommand{\s}{\mathrm{s}}
\newcommand{\Pa}{\mathrm{Pa}}

\newcommand{\Watt}{\mathrm{W}}
\newcommand{\Joule}{\mathrm{J}}

\begin{document}

\title{Giant Casimir non-equilibrium forces drive coil to globule transition in polymers}
\author{Himadri S. Samanta}\affiliation{Department of Chemistry, University of Texas at Austin, TX 78712}
\author{Mauro L. Mugnai}\affiliation{Department of Chemistry, University of Texas at Austin, TX 78712}
\author{T. R. Kirkpatrick}\affiliation{Institute For Physical Science and Technology, University of Maryland, College Park, MD 20742}
\author{D. Thirumalai} \affiliation{Department of Chemistry, University of Texas at Austin, TX 78712}
%\affiliation{Institute For Physical Science and Technology, University of Maryland, College Park, MD 20742}

%\begin{document}
%\maketitle

\begin{abstract}
We develop a theory to probe the effect of non-equilibrium fluctuation-induced forces on the size of a  polymer  confined between two horizontal thermally conductive plates subject to a constant temperature gradient, $\nabla T$. We assume that  (a) the solvent is good and (b) the distance between the plates is large so that in the absence of a thermal gradient the polymer is a coil whose size scales with the number of monomers as $N^{\nu}$, with $\nu \approx 0.6$.  We predict that above a critical temperature gradient, $\nabla T_c \sim N^{-\frac{5}{4}}$, favorable attractive monomer-monomer interaction due to Giant Casimir Force (GCF)  overcomes the chain conformational entropy, resulting  in a coil-globule transition. The long-ranged GCF-induced  interactions between monomers, arising from  thermal fluctuations in non-equilibrium steady state, depend on the thermodynamic properties of the fluid. Our predictions can be verified using light-scattering experiments with polymers, such as polystyrene or polyisoprene in organic solvents (neopentane) in which GCF is attractive.
\end{abstract}

\date{\today}
\maketitle
%\doublespacing

%\section{Introduction}
Interaction forces in nature are often caused by fluctuations of some physical entity in restricted geometries. 
 Well-known examples include the
Casimir interaction, which is the macroscopic manifestation of quantum fluctuations of the electromagnetic field. The Casimir interaction force, $f_{EM}$, arises due to the changes in the vacuum energy density, in the presence of two neutral perfectly conducting boundaries~\cite{Casimir48PKNA,Kardar99RMP}. These ideas were further generalized by Lifshitz for dielectric material characterized by frequency dependent dielectric permittivity~\cite{Lifshitz56SP,Dzyaloshinskii61AP}. 

Subsequently, Fisher and de Gennes remarked that a similar effect could emerge in condensed phases as well~\cite{Fisher78CRSACSB}. For confined critical systems, such as 
a fluid near the liquid-gas critical point, a binary liquid near the consolute point, or liquid $\text{He}^4$ near $\lambda$ transition, critical fluctuations of the order parameter generate long range forces between the confining walls, denoted by critical Casimir forces, $f_c$~\cite{Krech94}. 
A recent series of studies investigated the nature of the fluctuation-induced force generated between two parallel plates in a fluid with a temperature gradient $\nabla T$~\cite{Kirkpatrick13PRL,Kirkpatrick14PRE,Kirkpatrick16PRE}. This force results from the thermal fluctuations in a fluid in non-equilibrium (NE) steady state. Correlations of fluctuations in a NE fluid are generally longer range than other correlations, including those near an 
equilibrium critical point. For large distances, $f_{NE}>f_c>f_{EM}$~\cite{Kirkpatrick13PRL}. 

Temperature gradient-induced effects on colloids and polymers have been previously investigated in a variety of contexts. Examples include separation of macromolecules in organic solvents~\cite{Giddings93S}, crowding of nucleotides~\cite{Baaske07PNAS, Budin09JACS}, colloidal accumulation in micro-fluid~\cite{Jiang09PRL}, phase separation in a miscible polymer solution~\cite{Kumaki96PRL},  structural evolution in directional crystallization of polymers~\cite{Toda12M}, and thermophoresis~\cite{Piazza08JPCM}.~Here, we investigate the effect of the non-equilibrium giant Casimir force (GCF)~\cite{Kirkpatrick13PRL}, $f_{NE}$, on the size of a polymer. 
\begin{figure}[b]
  \includegraphics[width=0.38\textwidth]{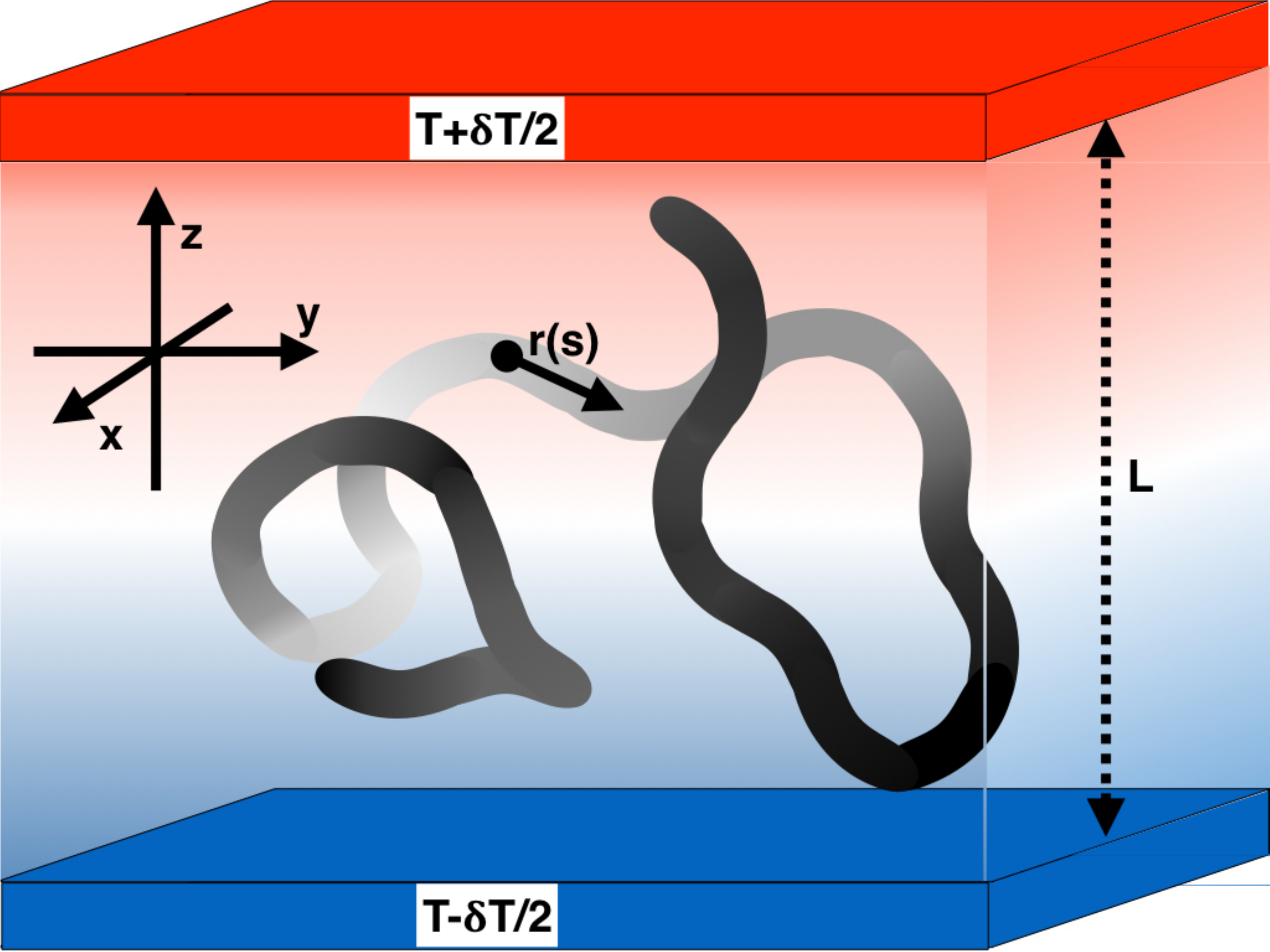}
\caption{A schematic of a homopolymer in a fluid (good solvent) confined between two parallel thermally conducting plates located at $z=0$ and $z=L$. 
The temperature difference between the plates is $\delta T$ and $\nabla T=\frac{\delta T}{L}$.}
	\label{chain}
\end{figure}
We develop a theory to assess the effects of GCF on a self-avoiding homopolymer with $N$ monomers of size $a_0$ confined between two parallel thermal conducting plates separated by a distance $L$ with a uniform temperature gradient, $\nabla T$ (Fig.~\ref{chain}). We assume that the fluid is a good solvent for the polymer, implying that the radius of gyration of the polymer is $R_g \sim a_0N^{\nu}$ with $\nu \approx 0.6$ if $\nabla T =0$; furthermore, we ignore the effect of confinement, which is to say that $\frac{L}{R_g} \gg 1$. In this setup the polymer behaves as a Flory random coil. %In the absence of $\nabla T$ collapse transition  is driven by the interplay of chain conformational entropy (dominating in the expanded coil state) and intra polymer interaction energy (dominating in the collapsed globular state). The swollen state is realized in good solvents (interaction between monomer and solvents is favorable), whereas in the collapsed state monomer-monomer interactions are preferred. The conformational change from a coil to a globule state in large polymers is akin to a phase transition~\cite{}. 
We show that the presence of the GCF induces an intramolecular attraction, which we predict to be sufficiently strong to overcome the conformational entropy of the polymer, thus inducing a genuine NE coil to globule transition. Such a transition, which is akin to a phase transition ~\cite{de1979scaling}, occurs in {\it equilibrium} ($\nabla T= 0$) only when the solvent quality is changed from good to poor. For a fixed $N$, we predict that there is a critical temperature gradient, $\nabla T_c$, above which the polymer undergoes a coil $\rightarrow$ globule transition. The critical temperature gradient at which the monomer-solvent energetics nearly compensates for the monomer-monomer interactions is defined as the NE $\Theta$-point. %scales as $\nabla T_c \approx N^{-\frac{5}{4}}$. 
%\section{Theory:}

Consider a polymer chain (Fig.~\ref{chain})~dissolved in a fluid confined between two horizontal thermal conducting plates, located at $z=0$ and $z=L$ with stationary temperature gradient $\nabla T$~(The local average temperature, $T(z)$, is a linear function of the coordinate $z$). We ignore the presence of convection in the fluid, which is achieved by setting the temperature of the upper plate at a larger value than the the temperature of the lower plate. 

We start with the Edwards Hamiltonian for a polymer chain,
\begin{equation}\label{hamiltonian}
\mathcal{H}_E=\frac{3k_B T}{2 a_0^2} \int\limits_0^N  \left(\frac{\partial {\bf r}}{\partial s}\right)^2 ds + k_B T  \mathcal{V}({\bf r}(s)),
\end{equation}
where, ${\bf r}(s)$ is the position of the monomer $s$ (Fig.~\ref{chain}), $a_0$ is the monomer size, and $N$ is the number of monomers. The first term in  Eq.~(\ref{hamiltonian}) accounts for chain connectivity, and the second term represents volume interactions given by $\mathcal{V}({\bf r}(s))=\frac{v}{(2\pi a_0^2)^{3/2}}\sum\limits_{s,s'=0}^{N}  
\text{exp}[{-\frac{({\bf r}(s)-{\bf r}(s'))^2}{2a_0^2}}]$. 
%\begin{equation}\label{HamiltV}
%\mathcal{V}({\bf r}(s))=\frac{v}{(2\pi a_0^2)^{3/2}}\sum\limits_{s,s'=0}^{N}  
%e^{-\frac{({\bf r}(s)-{\bf r}(s'))^2}{2a_0		^2}}-
%V( r_1 ,r_2 )%\label{HamiltV}
%\end{equation}
%We consider a chain whose size is smaller than the separation between plates, and therefore the plates have no effect on polymer~\cite{Corderio97JPF}. 
We consider a long polymer whose size, $R_g$, is much less than $L$. In this limit, it is known that $R_g \approx a_0 N^\nu$ with $\nu \approx 0.6 $ provided $v > 0$ 
i.e., the polymer is in a good solvent. If $L/R_g \ll 1$ then the polymer swells with $R_g\sim a_0 N^\nu$ with $\nu=3/4$~\cite{Corderio97JPF,DaoudJP77}, a situation not considered here.
In the presence of $\nabla T$, the interaction between two monomers at ${\bf r}_1 = {\bf r}(s_1)$ and ${\bf r}_2 = {\bf r}(s_2)$, is altered due to fluctuation-induced non-equilibrium forces.  The additional effective pair potential
due to GCF is $V({\bf r}_1,{\bf r}_2)$, which could be attractive or repulsive depending on the thermodynamic properties of the fluid. 

In the presence of $\nabla T$, the non-equilibrium fluctuations in fluids
are large and long ranged~\cite{Dorfman94ARPC}. The NE fluctuations arise due to the coupling between the heat and the viscous modes. The former is caused by  temperature fluctuations whereas the latter is due to velocity
fluctuations, which is accounted  by the convective term in the fluctuating-hydrodynamics equations~\cite{Croccolo16EPJE}. In a one component fluid, two sound modes are encountered, associated with pressure fluctuations, in addition to heat and viscous mode~\cite{Ortiz06,Chandrasekhar81}. 
In normal fluids, sound modes are fast propagating, while the heat mode is the slow diffusive mode. Therefore, to deal with slow diffusive temperature fluctuations, we may neglect pressure fluctuations.

For a quiescent fluid, in the presence of a uniform temperature gradient $\nabla T$, the non-equilibrium contribution to the intensity of the temperature fluctuations varies with the
wave  number $k$, diverging as $k^{-4}$ in the limit $k \rightarrow 0$, as predicted theoretically~\cite{Kirkpatrick82PRA,Law89JSP,Law89PRA,Belitz05RMP} and confirmed experimentally~\cite{Law90PRA, Segre92PRA, Segre93PRE,Li94PA, Vailati96PRL, Vailati97Nature, Takacs11PRL, Cerbino15SR}. As a consequence, the long-range NE fluctuations induce significant Casimir-like forces in fluids, in the presence of temperature gradient. Such a GCF should have a dramatic effect on the size of the polymer. 
 
The $L$-dependent NE fluctuation contribution to pressure, $P_{NE}(L)$, is given by~\cite{Kirkpatrick13PRL,Kirkpatrick14PRE,Kirkpatrick16PRE},
\begin{eqnarray}\label{free-energy}
&&P_{NE}(L)=k_B T ~A~L (\frac{\nabla T}{T})^2=k_B T \frac{A}{L} (\frac{\delta T}{T})^2, ~\text{with}, \\ \nonumber &&A=
\frac{C_P  T (\gamma-1) }{96 \pi D_T(\nu+D_T)} \left[ 1-\frac{1}{\alpha C_p}\left( \frac{\partial C_P}{\partial T} \right )_P +\frac{1}{\alpha^2} \left( \frac{\partial \alpha}{\partial T}  \right)_P \right]
\end{eqnarray}
where, $\delta T$ is the temperature difference between the two plates (Fig.~\ref{chain}), $C_P$ is the isobaric specific heat capacity, $D_T$ is the thermal diffusivity, $\gamma$ is the ratio of isobaric and isochoric heat capacities, and $\alpha$ is the thermal expansion coefficient.
%We note that the NE pressure depends on the vertical position $z$ (Fig.~\ref{chain}) and parametrically on the distance $L$ between the two horizontal plates.
For a fixed value of the temperature gradient, the NE pressure grows with increasing $L$. This anomalous behavior is a reflection of the very long-range spatial correlations in a fluid, in the presence of a temperature gradient. 
%The average fluctuation induced NE pressure is given by $\bar{P}_{NE}=\int^L_0 dz~P(z,L)=\frac{C_P k_B (\gamma-1) }{96 \pi D_T(\nu+D_T)} \left[ 1-\frac{1}{\alpha C_p}\left( \frac{\partial C_P}{\partial T} \right )_P +\frac{1}{\alpha^2} \left( \frac{\partial \alpha}{\partial T}  \right)_P \right] L (\frac{\nabla T}{T})^2$. 
For a fixed value of the temperature difference $\delta T$ between the plates, the giant NE Casimir pressure varies as $L^{-1}$, which is much longer ranged than the $f_{EM} \sim L^{-4}$ dependence  
 for electromagnetic Casimir forces, or a $f_c \sim L^{-3}$ dependence for the critical Casimir forces~\cite{Fisher78CRSACSB}. 

%To understand the physical origin of the NE Casimir pressure, we note that in general $\nabla T$ can cause normal stresses or pressures, if non-equilibrium thermodynamics is extended to include non-linear effects. The non-linear Onsagar-like cross effect, causing a non-equilibrium contribution to the pressure induced by a temperature gradient, is given by $P_{NE}=\kappa_{NL} (\nabla T)^2$, where $\kappa_{NL}$ is a nonlinear kinetic coefficient, commonly referred as a nonlinear Burnett coefficient~\cite{Wong78JCM}. $\kappa_{NL}$ is the sum of a bare molecular contribution, $\kappa_{NL}^{(0)}$, associated with short-range correlations, and a long-range fluctuating hydrodynamic contribution 
%$\kappa_{NL}^{(1)}L$ diverging as $L\rightarrow \infty$, $\kappa_{NL}=\kappa_{NL}^{(0)}+\kappa_{NL}^{(1)}L$~\cite{Kirkpatrick13PRL,Kirkpatrick14PRE,Ernst75JST, Brey83JCM}. The contribution of the bare term of Burnett coefficient to NE Casimir pressure is small compared to  $\kappa_{NL}^{(1)}L$.
%Hence, a NE Casimir pressure due to long range correlations is given by $P_{NE}(L)=\kappa_{NL}^{(1)}L (\nabla T)^2$, which has a similar form as the average NE Casimir pressure. We suggest that NE pressure due to long-rangecorrelations  is a direct consequence of the divergence of the corresponding nonlinear Burnett coefficient $\kappa_{NL}$. 

If a polymer is inserted between the plates then the average force between two monomers, which are at a distance $\mid z_1 -z_2 \mid$ apart, is given by ~\cite{Kirkpatrick16PRELM},
\begin{equation}\label{fe2}
 \mathcal{F}_{12} = -4 \pi a_0^2P_{NE}(L)\frac{|z_1 - z_2|}{L},
\end{equation}
where $a_0^2$ is the size of the monomer and $z_1$ and $z_2$ are the positions of the two monomers on polymer chain. We assume that $\nabla T$ acts  over a distance $|z_1 - z_2| $ such that $z_1 = L/2 - \delta/2$ and $z_2 = L/2 - \delta/2$ with $a_0 \ll \delta \ll L$.
The force can be attractive or repulsive depending on the nature of the fluid, which we discuss further below.

In order to calculate the $R_g$ of the polymer in the slit with $\nabla T$ we use an approximate method introduced by Edwards and Singh (ES)~\cite{Edwards79}.
The ES method is a variational type calculation that represents the exact Hamiltonian by a Gaussian chain. The effective monomer size in the variational Hamiltonian is determined as follows.
Consider a virtual chain without excluded volume interactions, with the radius of gyration $\langle R_{g}^{2} \rangle=N a^{2}/6$~\cite{Edwards79}, described by the Hamiltonian
\begin{equation}
\mathcal{H}_v=\frac{3k_B T}{2 a^2} \int\limits_0^N  \left(\frac{\partial {\bf r}}{\partial s}\right)^2 ds,
\end{equation}
where $a$ is the effective monomer size.
We split the deviation $\mathcal{W}$ between the virtual chain Hamiltonian and the exact Hamiltonian as,
\begin{equation}
%\vec{\chi}(s,t)={\bf r}(s,t)-{\bf r}_v (s,t),
\mathcal{H}-\mathcal{H}_v=k_BT\mathcal{W}=k_BT(\mathcal{W}_1+\mathcal{W}_2),
%&=&\mathcal{W}_1+\mathcal{W}_2
\end{equation}
where
\begin{eqnarray}\label{GCF}
&&\mathcal{W}_1=\frac{3}{2 }\left(\frac{1}{a_0^2}-\frac{1}{a^2}\right) \int\limits_0^N  \left(\frac{\partial {\bf r}}{\partial s}\right)^2 ds, \\ \nonumber
&&\mathcal{W}_2=\mathcal{V}({\bf r}(s)) + \sum\limits_{s,s'=0}^{N} V({\bf r}(s) - {\bf r}(s')).
\end{eqnarray}
The  term $V({\bf r}(s) - {\bf r}(s'))$ in Eq.~(\ref{GCF}) is the contribution from the GCF, and is obtained by integrating $ \mathcal{F}_{12}$ in Eq.~ (\ref{fe2}) with respect to the $z$-variable.
The radius of gyration is $R_g^2=\frac{1}{N} \int\limits_0^N \langle{\bf r}^2(s)\rangle ds$, with the average being
%\begin{equation}
	$\langle{\bf r}^2(s)\rangle=\frac{\int r^2 e^{-\mathcal{H}_v/k_BT}e^{\mathcal{-W}} \delta{\bf r}}{\int e^{-\mathcal{H}_v/k_BT}e^{\mathcal{-W}} \delta{\bf r}}=\frac{\langle{\bf r}^2(s)e^{\mathcal{-W}}\rangle_v}{\langle e^{\mathcal{-W}}\rangle_v}$,
%\end{equation}
where, $\langle \cdots \rangle_v$ denotes the average over $\mathcal{H}_v$.

Assuming that the deviation $\mathcal{W}$ is 'small', we can calculate the average $R_g$ to first order in $\mathcal{W}$. The result is, 
%\begin{equation}
$	\langle{\bf r}^2(s)\rangle \approx \frac{\langle{\bf r}^2(s)(1-\mathcal{W})\rangle_v}{\langle (1-\mathcal{W})\rangle_v} \approx \langle{\bf r}^2(s)(1-\mathcal{W})\rangle_v\langle (1+\mathcal{W})\rangle_v $,
%\end{equation}
leading to,
\begin{eqnarray}\label{rg}
&&\langle R_g^2\rangle=\frac{1}{N} \int\limits_0^N \langle{\bf r}^2(s)\rangle ds\\ \nonumber && = \frac{1}{N} \int\limits_0^N [\langle{\bf r}^2(s)\rangle_v + \langle{\bf r}^2(s)\rangle_v \langle\mathcal{W}\rangle_v -\langle{\bf r}^2(s)\mathcal{W}\rangle_v] ds.
\end{eqnarray}
If we choose the effective monomer size $a$ in $\mathcal{H}_v$ such that the first order correction (second and third terms in the second line of Eq.~(\ref{rg})) vanishes, then the size of the chain is, $\langle R_{g}^{2} \rangle=N a^{2}/6$. This is an estimate of the exact $\langle R_g^2 \rangle$, and is an approximation as we have neglected $\mathcal{W}^2$ and higher powers of $\mathcal{W}$. However, following the analysis in ~\cite{Edwards79} we can assume that inclusion of higher order terms merely renormalizes the coefficients of the dependence of $R_g$ on $N$ and $\nabla T$ without altering the essential qualitative results. Thus, in the ES theory, we find $a$ by setting the first order correction (second and third terms in the second line of  Eq.~(\ref{rg})) to zero.
%\begin{equation}\label{first}
 %\frac{1}{N} \int\limits_0^N [ \langle{\bf r}^2(s)\rangle_v \langle\mathcal{W}\rangle_v -\langle{\bf r}^2(s)\mathcal{W}\rangle_v] ds=0.
%\end{equation}
 The resulting equation should be solved in a self-consistent manner for $a$, and is given by~\cite{Edwards79}:
\begin{equation}
 \frac{1}{a_0^2}-\frac{1}{a^2}=
\frac{  \frac{1}{N}\int\limits_0^N [\langle{\bf r}^2(s)\rangle_v \langle\mathcal{W}_2\rangle_v -\langle{\bf r}^2(s)\mathcal{W}_2\rangle_v]ds}{ \frac{a^2}{N}\int_{0}^N ds \ \langle{\bf r}^2(s)\rangle_v}.
\end{equation}
Calculating the averages in the Fourier space (${\bf r}_n=\frac{1}{N}\int\limits_1^N \cos\left({ \frac{\pi n s}{N}}\right) {\bf r}(s) ds$; ${\bf r}(s)=2\sum\limits_{n =1}^{N}\cos\left({\frac{\pi n s}{N}}\right){\bf r}_n$; $R_g^2=2\sum\limits_n \langle|{{\bf r}_n}^2|\rangle$), we obtain
\begin{equation}\label{aaaa}
\frac{1}{a_0^2}-\frac{1}{a^2}=\sum\limits_{s,s'=0}^N  \left[
   v  \frac{C^{ss'}_{1}}{(C^{ss'}_{})^{5/2}}
- A \left(\frac{\nabla T}{T}\right)^2 C^{ss'}_{2}\right]
\end{equation}
where, $C^{ss'}_{1}= \frac{\sqrt{2}N}{3 a^5\pi^{5/2}\sum\limits_{n=1}^N\frac{1}{n^2}}\sum\limits_{n=1}^{N}2\frac{1-\cos[n \pi(s-s')/N]}{n^4}$, $C^{ss'}=\frac{2N}{3 \pi^2}\sum\limits_{n=1}^{N}\frac{1-\cos[n \pi(s-s')/N]}{n^2}+\frac{a_0^2 }{a^2 }$, $C^{ss'}_{2}=\frac{4a_0^2 N}{{9\pi \sum\limits_{n=1}^N\frac{1}{n^2} }}\sum\limits_{n=1}^{N}2\frac{1-\cos[n \pi(s-s')/N]}{n^4}$, and $v=\frac{4}{3}\pi a_0^3$.

The best estimate for the effective monomer size $a$ can be obtained using Eq.~(\ref{aaaa}).  The $\Theta$-point signals the transition from a  coil to a globule. We use the definition of the $\Theta$-point to assess the condition for collapse in terms of temperature gradient, instead of solving the complicated Eq.~(\ref{aaaa}) numerically. The volume interactions are on the right hand side of Eq.~(\ref{aaaa}). At the $\Theta$-point, the $v$-term should exactly balance the GCF term. Since at the $\Theta$-point the dimensions of the chain is ideal, implying  $a=a_0$, we can substitute this value for $a$ in the $v$- and the GCF terms, and equate the two . The result yields an expression for the $\Theta$-point as a function of $\nabla T$.
Thus, from Eq.~(\ref{aaaa}) the temperature gradient at which two body repulsion ($v$-term) equals two body interaction (Casimir-term)  is,
\begin{equation}
	\Big(\frac{\nabla T_c}{T}\Big)^2=\frac{ \sum\limits_{s,s'=0}^N  v \frac{C^{ss'}_{1}}{(C^{ss'}_{})^{5/2}}}
     {A\sum\limits_{s,s'=0}^N  C^{ss'}_{2}}.
\label{deltaT}
\end{equation}
The numerator in Eq.~(\ref{deltaT}) is a consequence of chain connectivity, and the denominator encodes the fluctuation induced effect, determining the extent to which the sizes in extended states change with temperature gradient.
Clearly, $\nabla T_c$ is determined by the fluid properties through $A$.
The results in Eq.~(\ref{aaaa}) can be used to obtain the dependence of $\nabla T_c$ on $N$. Scaling  $n $ by $N$, it can be shown that $C^{ss'}_{1}\sim N$. Similarly, $C^{ss'}_2 \sim N$ and $C^{ss'} \sim N$. From these  results it follows that 
\begin{equation}
\nabla T_c \sim  N^{-\frac{5}{4}}.
\end{equation} 

\begin{figure}[t]
	\includegraphics[width=0.43\textwidth]{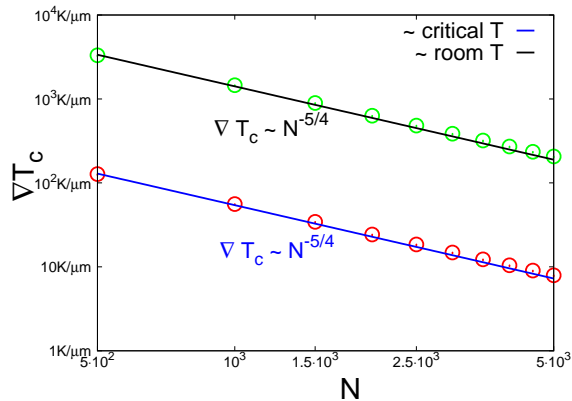}
       %\begin{small}
	\caption{ The critical temperature gradient $\nabla T_c$ decreases with $N$ as $N^{-5/4}$. Red and green circles are the predicted values for $\nabla T_c$ for neopentane fluid at temperature 433K and 300K respectively  for different polymer chains. The calculations are done for polystyrene for which $a_0$, the monomer size, is $\approx$ 17\AA~\cite{Suzuki13JCP}.}
	%`\end{small}
	\label{dtc}
\end{figure}

\begin{figure}[!tbp]
  \centering
  \subfloat{\includegraphics[width=0.46\textwidth]{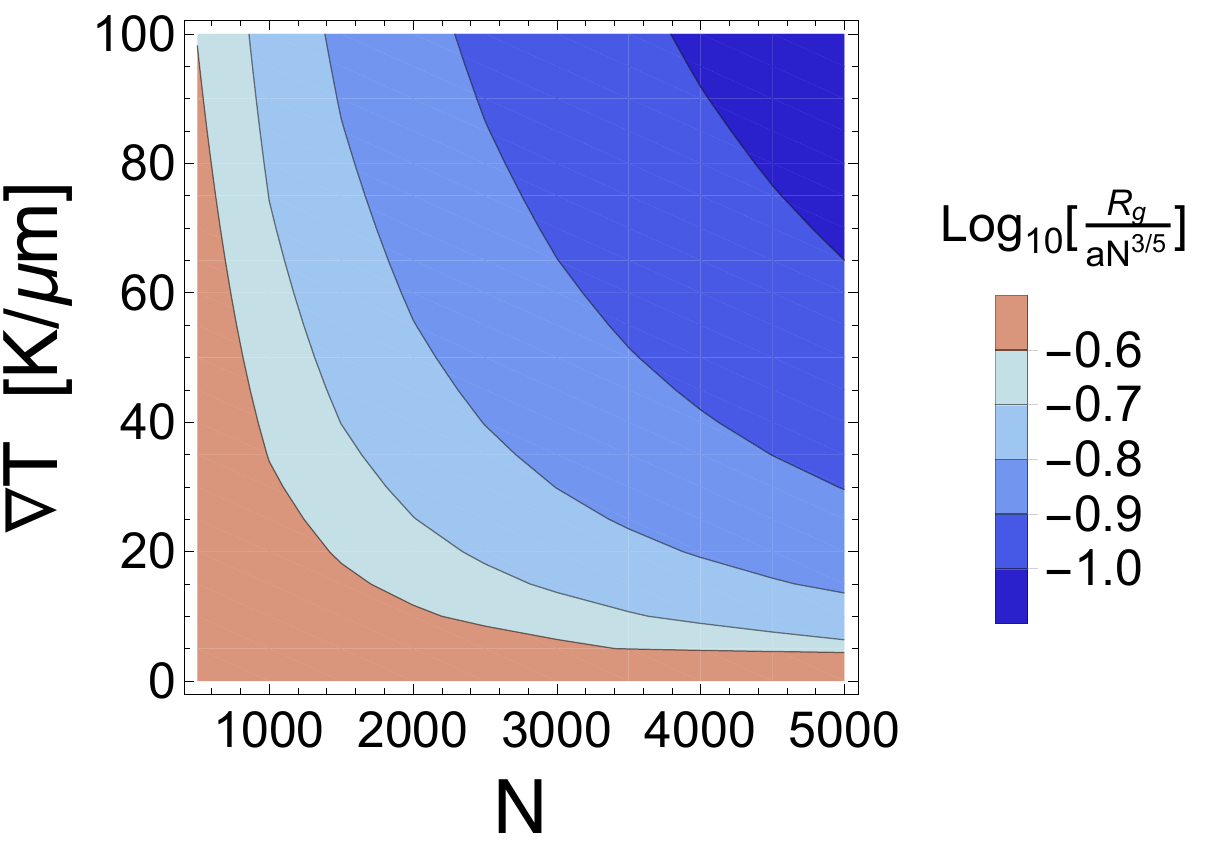}\label{fig:f1}}
  \hfill
  \vspace*{0.6cm}
  \subfloat{\includegraphics[width=0.46\textwidth]{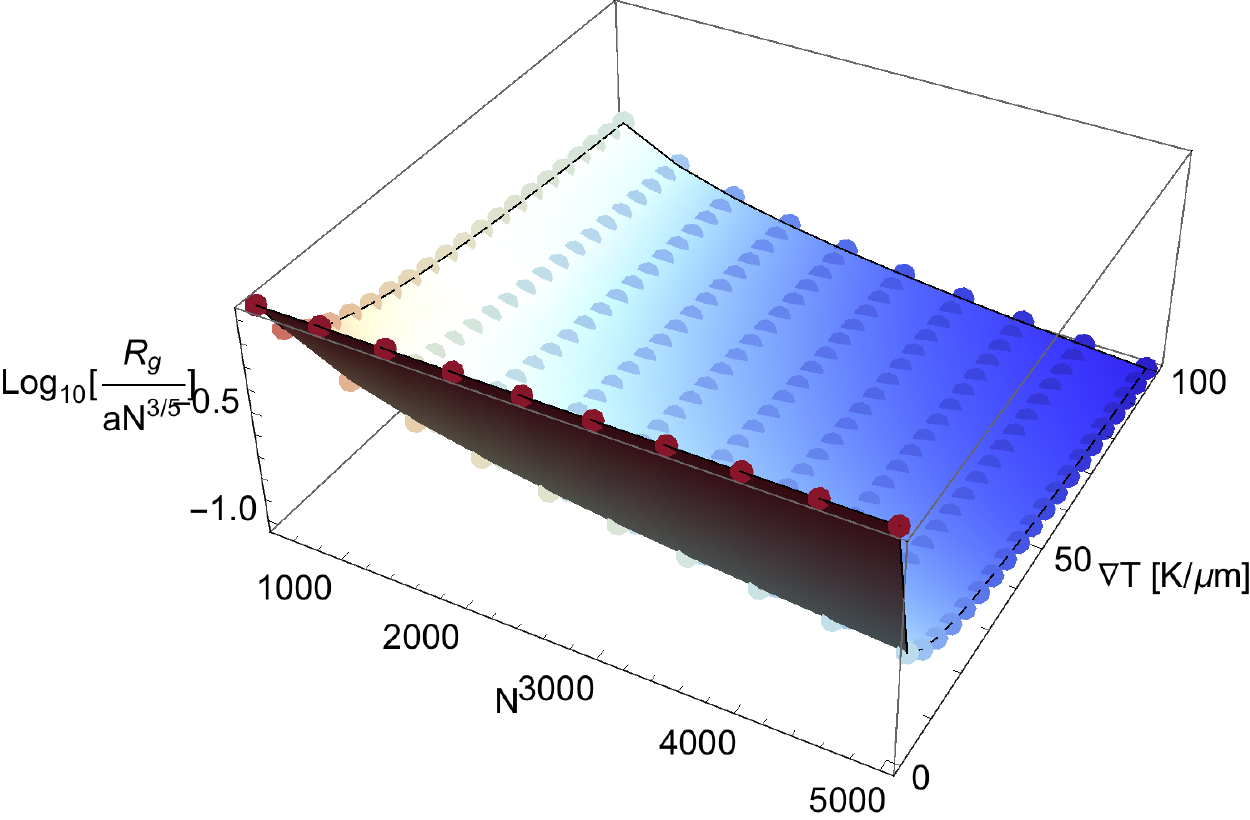}\label{fig:f2}}
  \caption{ The top panel shows that the radius of gyration ($R_g$) for ten different chain lengths (from N= 500 to N=5000).  decreases continuously with increasing temperature gradient. The bottom panel is a three dimensional plot displaying $R_g$ as a function of $N$ and $\nabla T$. These figures show significant contraction of the chain over the span of experimentally realizable values of the temperature gradient.}
  \label{RG}
\end{figure}

To estimate the critical temperature gradient $\nabla T_c$, implied by Eq.~(\ref{deltaT}), we consider the example of fluid neopentane, for which accurate light scattering experiments of the 
non-equilibrium temperature fluctuations are available~\cite{Lemmon06JCED}. 
Using available data for the thermodynamic and transport properties for neopentane [described in the Supplementary Information (SI)~\cite{Lemmon06JCED}], we calculated the numerical value of temperature gradient $\nabla T$ using Eq.~(\ref{deltaT}) at the 
$\Theta$ transition.~As an example consider the parameters given in Table~S1 in the SI~\cite{Lemmon06JCED}. The predicted value (Eq. 10) for $\nabla T_c$ near the critical temperature for neopentane  ranges from $\approx (35-8)~K \cdot {\mu m}^{-1}$ for $N$ between 1500 to 5000 (see the red line in Fig.~\ref{dtc}). This falls within the  experimentally achievable range of temperature gradient~\cite{Talbot17NC}. 
 The GCF term is negative for the fluid neopentane, implying that a polymer for which neopentane is a good solvent (such as polyisoprene or polystyrene) is predicted to undergo a coil to globule transition when $\nabla T$ exceeds $\nabla T_c$.~Fig.~\ref{RG} shows that the size of the chain ($R_g$) decreases continuously with increasing $\nabla T$, implying non-equilibrium fluctuations driven collapse of a polymer chain in a fluid.
The GCF term is very sensitive to the thermodynamic properties of the relevant fluid. 
For example, for toluene, the GCF is positive for temperature below 310K at 26 MPa pressure, implying  fluctuation-induced interactions between monomers are repulsive (see the SI for the values for toluene). In this case, the chain is  in a good solvent even if $\nabla T\ne 0$, and we predict that there ought to be no coil-globule transition for any value of $\nabla T$. Above 310K temperature, GCF term is negative, implying fluctuation induced interactions between monomers are attractive and the coil-globule transition would occur when $\nabla T$ exceeds $\nabla T_c$. These spectacularly contrasting predictions can be verified using currently available techniques using standard polymers (polystyrene or polyisoprene) in organic solvents. %From Eq.~(\ref{deltaT}), we observe that $\nabla T_c$  decreases with increasing $N$ as $1/N^{2.35}$,  exhibiting a complex $N$ dependence. 
%We evaluated the radius of gyration ($R_g$) for three different chain lengths (N= 500, 1000 and 1500). 
%Fig.~\ref{RG} shows clearly that the size of the chain ($R_g$) decreases continuously with increasing $\nabla T$, implying non-equilibrium fluctuations driven collapse of a polymer chain in a fluid, at non-equilibrium steady states.

In summary, we predict that the non-equilibrium giant Casimir force between monomers in a homopolymer could induce a coil to globule transition in a polymer that exists in  a swollen random coil state in the absence of the temperature gradient.
The interactions leading to the predicted dramatic transition is due to attractive long ranged interactions between the monomers due thermal fluctuations in the NE steady state. The fluctuation induced force could be attractive or repulsive depending on the thermodynamic properties of the relevant fluid. 
Our theory predicts that  above a critical value of the temperature gradient monomer-monomer attractive interactions will overcome the chain conformational entropy, inducing a coil to globule transition in certain solvents. The easiest experiment to imagine is a polymer in a custom-made cell in a fluid with the system initially in thermal equilibrium. A vertical temperature difference $\Delta T$ across the cell is then imposed. Assuming the temperature gradient in the solvent is quickly established, we predict that the polymer will respond to it and collapse or not according to our theoretical considerations. We should emphasize that because our predictions suggest a large change in the size of the polymer the effects are far from being subtle. Consequently, we envision that our predictions are  amenable to test by standard light scattering experiments.
 Finally we note that a polymer in liquid mixtures can also be profoundly affected by concentration fluctuations. In this case, a temperature gradient induces long-range concentration fluctuations through the Soret effect.  We hope to report the consequences of the Soret effect on the size of a polymer elsewhere.   
    
\section{Acknowledgements}
This work was supported by the National Science Foundation under Grant Nos. DMR-1401449 and CHE-16-36424. DT acknowledges additional support from the Collie-Welch Regents Chair (F-0019).
%\bibliographystyle{unsrt}
%\bibliography{SI}

\begin{thebibliography}{40}%
\makeatletter
\providecommand \@ifxundefined [1]{%
 \@ifx{#1\undefined}
}%
\providecommand \@ifnum [1]{%
 \ifnum #1\expandafter \@firstoftwo
 \else \expandafter \@secondoftwo
 \fi
}%
\providecommand \@ifx [1]{%
 \ifx #1\expandafter \@firstoftwo
 \else \expandafter \@secondoftwo
 \fi
}%
\providecommand \natexlab [1]{#1}%
\providecommand \enquote  [1]{``#1''}%
\providecommand \bibnamefont  [1]{#1}%
\providecommand \bibfnamefont [1]{#1}%
\providecommand \citenamefont [1]{#1}%
\providecommand \href@noop [0]{\@secondoftwo}%
\providecommand \href [0]{\begingroup \@sanitize@url \@href}%
\providecommand \@href[1]{\@@startlink{#1}\@@href}%
\providecommand \@@href[1]{\endgroup#1\@@endlink}%
\providecommand \@sanitize@url [0]{\catcode `\\12\catcode `\$12\catcode
  `\&12\catcode `\#12\catcode `\^12\catcode `\_12\catcode `\%12\relax}%
\providecommand \@@startlink[1]{}%
\providecommand \@@endlink[0]{}%
\providecommand \url  [0]{\begingroup\@sanitize@url \@url }%
\providecommand \@url [1]{\endgroup\@href {#1}{\urlprefix }}%
\providecommand \urlprefix  [0]{URL }%
\providecommand \Eprint [0]{\href }%
\providecommand \doibase [0]{http://dx.doi.org/}%
\providecommand \selectlanguage [0]{\@gobble}%
\providecommand \bibinfo  [0]{\@secondoftwo}%
\providecommand \bibfield  [0]{\@secondoftwo}%
\providecommand \translation [1]{[#1]}%
\providecommand \BibitemOpen [0]{}%
\providecommand \bibitemStop [0]{}%
\providecommand \bibitemNoStop [0]{.\EOS\space}%
\providecommand \EOS [0]{\spacefactor3000\relax}%
\providecommand \BibitemShut  [1]{\csname bibitem#1\endcsname}%
\let\auto@bib@innerbib\@empty
%</preamble>
\bibitem [{\citenamefont {Casimir}(1948)}]{Casimir48PKNA}%
  \BibitemOpen
  \bibfield  {author} {\bibinfo {author} {\bibfnamefont {H.~B.~G.}\
  \bibnamefont {Casimir}},\ }\href@noop {} {\bibfield  {journal} {\bibinfo
  {journal} {Proc. K. Ned. Akad. Wet.}\ }\textbf {\bibinfo {volume} {51}},\
  \bibinfo {pages} {793} (\bibinfo {year} {1948})}\BibitemShut {NoStop}%
\bibitem [{\citenamefont {Kardar}\ and\ \citenamefont
  {Golestanian}(1999)}]{Kardar99RMP}%
  \BibitemOpen
  \bibfield  {author} {\bibinfo {author} {\bibfnamefont {M.}~\bibnamefont
  {Kardar}}\ and\ \bibinfo {author} {\bibfnamefont {R.}~\bibnamefont
  {Golestanian}},\ }\href@noop {} {\bibfield  {journal} {\bibinfo  {journal}
  {Rev. Mod. Phys.}\ }\textbf {\bibinfo {volume} {71}},\ \bibinfo {pages}
  {1233} (\bibinfo {year} {1999})}\BibitemShut {NoStop}%
\bibitem [{\citenamefont {Lifshitz}(1956)}]{Lifshitz56SP}%
  \BibitemOpen
  \bibfield  {author} {\bibinfo {author} {\bibfnamefont {E.~M.}\ \bibnamefont
  {Lifshitz}},\ }\href@noop {} {\bibfield  {journal} {\bibinfo  {journal} {Sov.
  Phys. JETP}\ }\textbf {\bibinfo {volume} {2}},\ \bibinfo {pages} {73}
  (\bibinfo {year} {1956})}\BibitemShut {NoStop}%
\bibitem [{\citenamefont {Dzyaloshinskii}\ \emph {et~al.}(1961)\citenamefont
  {Dzyaloshinskii}, \citenamefont {Lifshitz},\ and\ \citenamefont
  {Pitaevskii}}]{Dzyaloshinskii61AP}%
  \BibitemOpen
  \bibfield  {author} {\bibinfo {author} {\bibfnamefont {I.}~\bibnamefont
  {Dzyaloshinskii}}, \bibinfo {author} {\bibfnamefont {E.}~\bibnamefont
  {Lifshitz}}, \ and\ \bibinfo {author} {\bibfnamefont {L.}~\bibnamefont
  {Pitaevskii}},\ }\href@noop {} {\bibfield  {journal} {\bibinfo  {journal}
  {Adv. Phys.}\ }\textbf {\bibinfo {volume} {10}},\ \bibinfo {pages} {165}
  (\bibinfo {year} {1961})}\BibitemShut {NoStop}%
\bibitem [{\citenamefont {Fisher}\ and\ \citenamefont
  {de~Gennes}(1978)}]{Fisher78CRSACSB}%
  \BibitemOpen
  \bibfield  {author} {\bibinfo {author} {\bibfnamefont {M.}~\bibnamefont
  {Fisher}}\ and\ \bibinfo {author} {\bibfnamefont {P.~G.}\ \bibnamefont
  {de~Gennes}},\ }\href@noop {} {\bibfield  {journal} {\bibinfo  {journal} {C.
  R. Seances Acad. Sci. B}\ }\textbf {\bibinfo {volume} {287}},\ \bibinfo
  {pages} {207} (\bibinfo {year} {1978})}\BibitemShut {NoStop}%
\bibitem [{\citenamefont {Krech}(1994)}]{Krech94}%
  \BibitemOpen
  \bibfield  {author} {\bibinfo {author} {\bibfnamefont {M.}~\bibnamefont
  {Krech}},\ }\href@noop {} {\emph {\bibinfo {title} {The Casimir Effect in
  Critical Systems}}}\ (\bibinfo  {publisher} {World Scientific, Singapore},\
  \bibinfo {year} {1994})\BibitemShut {NoStop}%
\bibitem [{\citenamefont {Kirkpatrick}\ \emph {et~al.}(2013)\citenamefont
  {Kirkpatrick}, \citenamefont {Ortiz~de Z\'arate},\ and\ \citenamefont
  {Sengers}}]{Kirkpatrick13PRL}%
  \BibitemOpen
  \bibfield  {author} {\bibinfo {author} {\bibfnamefont {T.~R.}\ \bibnamefont
  {Kirkpatrick}}, \bibinfo {author} {\bibfnamefont {J.~M.}\ \bibnamefont
  {Ortiz~de Z\'arate}}, \ and\ \bibinfo {author} {\bibfnamefont {J.~V.}\
  \bibnamefont {Sengers}},\ }\href@noop {} {\bibfield  {journal} {\bibinfo
  {journal} {{Phys. Rev. Lett.}}\ }\textbf {\bibinfo {volume} {110}},\ \bibinfo
  {pages} {235902} (\bibinfo {year} {2013})}\BibitemShut {NoStop}%
\bibitem [{\citenamefont {Kirkpatrick}\ \emph {et~al.}(2014)\citenamefont
  {Kirkpatrick}, \citenamefont {Ortiz~de Z\'arate},\ and\ \citenamefont
  {Sengers}}]{Kirkpatrick14PRE}%
  \BibitemOpen
  \bibfield  {author} {\bibinfo {author} {\bibfnamefont {T.~R.}\ \bibnamefont
  {Kirkpatrick}}, \bibinfo {author} {\bibfnamefont {J.~M.}\ \bibnamefont
  {Ortiz~de Z\'arate}}, \ and\ \bibinfo {author} {\bibfnamefont {J.~V.}\
  \bibnamefont {Sengers}},\ }\href@noop {} {\bibfield  {journal} {\bibinfo
  {journal} {Phys. Rev. E}\ }\textbf {\bibinfo {volume} {89}},\ \bibinfo
  {pages} {022145} (\bibinfo {year} {2014})}\BibitemShut {NoStop}%
\bibitem [{\citenamefont {Kirkpatrick}\ \emph
  {et~al.}(2016{\natexlab{a}})\citenamefont {Kirkpatrick}, \citenamefont
  {Ortiz~de Z\'arate},\ and\ \citenamefont {Sengers}}]{Kirkpatrick16PRE}%
  \BibitemOpen
  \bibfield  {author} {\bibinfo {author} {\bibfnamefont {T.~R.}\ \bibnamefont
  {Kirkpatrick}}, \bibinfo {author} {\bibfnamefont {J.~M.}\ \bibnamefont
  {Ortiz~de Z\'arate}}, \ and\ \bibinfo {author} {\bibfnamefont {J.~V.}\
  \bibnamefont {Sengers}},\ }\href@noop {} {\bibfield  {journal} {\bibinfo
  {journal} {Phys. Rev. E}\ }\textbf {\bibinfo {volume} {93}},\ \bibinfo
  {pages} {012148} (\bibinfo {year} {2016}{\natexlab{a}})}\BibitemShut
  {NoStop}%
\bibitem [{\citenamefont {Giddings}(1993)}]{Giddings93S}%
  \BibitemOpen
  \bibfield  {author} {\bibinfo {author} {\bibfnamefont {J.~C.}\ \bibnamefont
  {Giddings}},\ }\href@noop {} {\bibfield  {journal} {\bibinfo  {journal}
  {Science}\ }\textbf {\bibinfo {volume} {260}},\ \bibinfo {pages} {1456}
  (\bibinfo {year} {1993})}\BibitemShut {NoStop}%
\bibitem [{\citenamefont {Baaske}\ \emph {et~al.}(2007)\citenamefont {Baaske},
  \citenamefont {Weinert}, \citenamefont {Duhr}, \citenamefont {K.~Lemke},\
  and\ \citenamefont {Braun}}]{Baaske07PNAS}%
  \BibitemOpen
  \bibfield  {author} {\bibinfo {author} {\bibfnamefont {P.}~\bibnamefont
  {Baaske}}, \bibinfo {author} {\bibfnamefont {F.}~\bibnamefont {Weinert}},
  \bibinfo {author} {\bibfnamefont {S.}~\bibnamefont {Duhr}}, \bibinfo {author}
  {\bibfnamefont {M.~R.}\ \bibnamefont {K.~Lemke}}, \ and\ \bibinfo {author}
  {\bibfnamefont {D.}~\bibnamefont {Braun}},\ }\href@noop {} {\bibfield
  {journal} {\bibinfo  {journal} {{Proc. Natl. Acad. Sci.}}\ }\textbf {\bibinfo
  {volume} {104}},\ \bibinfo {pages} {9346} (\bibinfo {year}
  {2007})}\BibitemShut {NoStop}%
\bibitem [{\citenamefont {Budin}\ \emph {et~al.}(2009)\citenamefont {Budin},
  \citenamefont {Bruckner},\ and\ \citenamefont {Szostak}}]{Budin09JACS}%
  \BibitemOpen
  \bibfield  {author} {\bibinfo {author} {\bibfnamefont {I.}~\bibnamefont
  {Budin}}, \bibinfo {author} {\bibfnamefont {R.}~\bibnamefont {Bruckner}}, \
  and\ \bibinfo {author} {\bibfnamefont {J.}~\bibnamefont {Szostak}},\
  }\href@noop {} {\bibfield  {journal} {\bibinfo  {journal} {J. Am. Chem.
  Soc.}\ }\textbf {\bibinfo {volume} {131}},\ \bibinfo {pages} {9628} (\bibinfo
  {year} {2009})}\BibitemShut {NoStop}%
\bibitem [{\citenamefont {Jiang}\ \emph {et~al.}(2009)\citenamefont {Jiang},
  \citenamefont {Wada}, \citenamefont {Yoshinaga},\ and\ \citenamefont
  {Sano}}]{Jiang09PRL}%
  \BibitemOpen
  \bibfield  {author} {\bibinfo {author} {\bibfnamefont {H.-R.}\ \bibnamefont
  {Jiang}}, \bibinfo {author} {\bibfnamefont {H.}~\bibnamefont {Wada}},
  \bibinfo {author} {\bibfnamefont {N.}~\bibnamefont {Yoshinaga}}, \ and\
  \bibinfo {author} {\bibfnamefont {M.}~\bibnamefont {Sano}},\ }\href@noop {}
  {\bibfield  {journal} {\bibinfo  {journal} {{Phys. Rev. Lett.}}\ }\textbf
  {\bibinfo {volume} {102}},\ \bibinfo {pages} {208301} (\bibinfo {year}
  {2009})}\BibitemShut {NoStop}%
\bibitem [{\citenamefont {Kumaki}\ \emph {et~al.}(1996)\citenamefont {Kumaki},
  \citenamefont {Hashimoto},\ and\ \citenamefont {Granick}}]{Kumaki96PRL}%
  \BibitemOpen
  \bibfield  {author} {\bibinfo {author} {\bibfnamefont {J.}~\bibnamefont
  {Kumaki}}, \bibinfo {author} {\bibfnamefont {T.}~\bibnamefont {Hashimoto}}, \
  and\ \bibinfo {author} {\bibfnamefont {S.}~\bibnamefont {Granick}},\
  }\href@noop {} {\bibfield  {journal} {\bibinfo  {journal} {{Phys. Rev.
  Lett.}}\ }\textbf {\bibinfo {volume} {77}},\ \bibinfo {pages} {1990}
  (\bibinfo {year} {1996})}\BibitemShut {NoStop}%
\bibitem [{\citenamefont {Toda}\ \emph {et~al.}(2012)\citenamefont {Toda},
  \citenamefont {Taguchi},\ and\ \citenamefont {Kajioka}}]{Toda12M}%
  \BibitemOpen
  \bibfield  {author} {\bibinfo {author} {\bibfnamefont {A.}~\bibnamefont
  {Toda}}, \bibinfo {author} {\bibfnamefont {K.}~\bibnamefont {Taguchi}}, \
  and\ \bibinfo {author} {\bibfnamefont {H.}~\bibnamefont {Kajioka}},\
  }\href@noop {} {\bibfield  {journal} {\bibinfo  {journal} {Macromolecules}\
  }\textbf {\bibinfo {volume} {45}},\ \bibinfo {pages} {852} (\bibinfo {year}
  {2012})}\BibitemShut {NoStop}%
\bibitem [{\citenamefont {Piazza}\ and\ \citenamefont
  {Parola}(2008)}]{Piazza08JPCM}%
  \BibitemOpen
  \bibfield  {author} {\bibinfo {author} {\bibfnamefont {R.}~\bibnamefont
  {Piazza}}\ and\ \bibinfo {author} {\bibfnamefont {A.}~\bibnamefont
  {Parola}},\ }\href@noop {} {\bibfield  {journal} {\bibinfo  {journal} {J.
  Phys.: Condens. Matter}\ }\textbf {\bibinfo {volume} {20}},\ \bibinfo {pages}
  {153102} (\bibinfo {year} {2008})}\BibitemShut {NoStop}%
\bibitem [{\citenamefont {de~Gennes}(1979)}]{de1979scaling}%
  \BibitemOpen
  \bibfield  {author} {\bibinfo {author} {\bibfnamefont {P.~G.}\ \bibnamefont
  {de~Gennes}},\ }\href@noop {} {\emph {\bibinfo {title} {Scaling Concepts in
  Polymer Physics}}}\ (\bibinfo  {publisher} {Cornell University Press},\
  \bibinfo {year} {1979})\BibitemShut {NoStop}%
\bibitem [{\citenamefont {Corderio}\ \emph {et~al.}(1997)\citenamefont
  {Corderio}, \citenamefont {Molisana},\ and\ \citenamefont
  {Thirumalai}}]{Corderio97JPF}%
  \BibitemOpen
  \bibfield  {author} {\bibinfo {author} {\bibfnamefont {C.}~\bibnamefont
  {Corderio}}, \bibinfo {author} {\bibfnamefont {M.}~\bibnamefont {Molisana}},
  \ and\ \bibinfo {author} {\bibfnamefont {D.}~\bibnamefont {Thirumalai}},\
  }\href@noop {} {\bibfield  {journal} {\bibinfo  {journal} {J. Phys. II
  France}\ }\textbf {\bibinfo {volume} {7}},\ \bibinfo {pages} {433} (\bibinfo
  {year} {1997})}\BibitemShut {NoStop}%
\bibitem [{\citenamefont {Daoud}\ and\ \citenamefont
  {De~Gennes}(1977)}]{DaoudJP77}%
  \BibitemOpen
  \bibfield  {author} {\bibinfo {author} {\bibfnamefont {M.}~\bibnamefont
  {Daoud}}\ and\ \bibinfo {author} {\bibfnamefont {P.-G.}\ \bibnamefont
  {De~Gennes}},\ }\href@noop {} {\bibfield  {journal} {\bibinfo  {journal} {J.
  Phys. (Paris)}\ }\textbf {\bibinfo {volume} {38(1)}},\ \bibinfo {pages} {85}
  (\bibinfo {year} {1977})}\BibitemShut {NoStop}%
\bibitem [{\citenamefont {Dorfman}\ \emph {et~al.}(1994)\citenamefont
  {Dorfman}, \citenamefont {Kirkpatrick},\ and\ \citenamefont
  {Sengers}}]{Dorfman94ARPC}%
  \BibitemOpen
  \bibfield  {author} {\bibinfo {author} {\bibfnamefont {J.~R.}\ \bibnamefont
  {Dorfman}}, \bibinfo {author} {\bibfnamefont {T.~R.}\ \bibnamefont
  {Kirkpatrick}}, \ and\ \bibinfo {author} {\bibfnamefont {J.~V.}\ \bibnamefont
  {Sengers}},\ }\href@noop {} {\bibfield  {journal} {\bibinfo  {journal} {Annu.
  Rev. Phys. Chem.}\ }\textbf {\bibinfo {volume} {45}},\ \bibinfo {pages} {213}
  (\bibinfo {year} {1994})}\BibitemShut {NoStop}%
\bibitem [{\citenamefont {Croccolo}\ \emph {et~al.}(2016)\citenamefont
  {Croccolo}, \citenamefont {Ortiz~de Z\'arate},\ and\ \citenamefont
  {Sengers}}]{Croccolo16EPJE}%
  \BibitemOpen
  \bibfield  {author} {\bibinfo {author} {\bibfnamefont {F.}~\bibnamefont
  {Croccolo}}, \bibinfo {author} {\bibfnamefont {J.~M.}\ \bibnamefont {Ortiz~de
  Z\'arate}}, \ and\ \bibinfo {author} {\bibfnamefont {J.}~\bibnamefont
  {Sengers}},\ }\href@noop {} {\bibfield  {journal} {\bibinfo  {journal} {Eur.
  Phys. J. E}\ }\textbf {\bibinfo {volume} {39}},\ \bibinfo {pages} {125}
  (\bibinfo {year} {2016})}\BibitemShut {NoStop}%
\bibitem [{\citenamefont {Ortiz~de Z\'arate}\ and\ \citenamefont
  {Sengers}(2006)}]{Ortiz06}%
  \BibitemOpen
  \bibfield  {author} {\bibinfo {author} {\bibfnamefont {J.~M.}\ \bibnamefont
  {Ortiz~de Z\'arate}}\ and\ \bibinfo {author} {\bibfnamefont {J.~V.}\
  \bibnamefont {Sengers}},\ }\href@noop {} {\emph {\bibinfo {title}
  {Hydrodynamic Fluctuations in Fluids and Fluid Mixtures}}}\ (\bibinfo
  {publisher} {Elsevier, Amsterdam},\ \bibinfo {year} {2006})\BibitemShut
  {NoStop}%
\bibitem [{\citenamefont {Chandrasekhar}(1981)}]{Chandrasekhar81}%
  \BibitemOpen
  \bibfield  {author} {\bibinfo {author} {\bibfnamefont {S.}~\bibnamefont
  {Chandrasekhar}},\ }\href@noop {} {\emph {\bibinfo {title} {Hydrodynamic and
  Hydromagnetic Stability}}}\ (\bibinfo  {publisher} {Oxford University
  Press/Dover, Oxford},\ \bibinfo {year} {1981})\BibitemShut {NoStop}%
\bibitem [{\citenamefont {Kirkpatrick}\ \emph {et~al.}(1982)\citenamefont
  {Kirkpatrick}, \citenamefont {Cohen},\ and\ \citenamefont
  {Dorfman}}]{Kirkpatrick82PRA}%
  \BibitemOpen
  \bibfield  {author} {\bibinfo {author} {\bibfnamefont {T.~R.}\ \bibnamefont
  {Kirkpatrick}}, \bibinfo {author} {\bibfnamefont {E.~G.~D.}\ \bibnamefont
  {Cohen}}, \ and\ \bibinfo {author} {\bibfnamefont {J.~R.}\ \bibnamefont
  {Dorfman}},\ }\href@noop {} {\bibfield  {journal} {\bibinfo  {journal} {Phys.
  Rev. A}\ }\textbf {\bibinfo {volume} {26}},\ \bibinfo {pages} {995} (\bibinfo
  {year} {1982})}\BibitemShut {NoStop}%
\bibitem [{\citenamefont {Law}\ and\ \citenamefont {Sengers}(1989)}]{Law89JSP}%
  \BibitemOpen
  \bibfield  {author} {\bibinfo {author} {\bibfnamefont {B.~M.}\ \bibnamefont
  {Law}}\ and\ \bibinfo {author} {\bibfnamefont {J.~V.}\ \bibnamefont
  {Sengers}},\ }\href@noop {} {\bibfield  {journal} {\bibinfo  {journal} {J.
  Stat. Phys.}\ }\textbf {\bibinfo {volume} {57}},\ \bibinfo {pages} {531}
  (\bibinfo {year} {1989})}\BibitemShut {NoStop}%
\bibitem [{\citenamefont {Law}\ and\ \citenamefont
  {Nieuwoudt}(1989)}]{Law89PRA}%
  \BibitemOpen
  \bibfield  {author} {\bibinfo {author} {\bibfnamefont {B.~M.}\ \bibnamefont
  {Law}}\ and\ \bibinfo {author} {\bibfnamefont {J.~C.}\ \bibnamefont
  {Nieuwoudt}},\ }\href@noop {} {\bibfield  {journal} {\bibinfo  {journal}
  {Phys. Rev. A}\ }\textbf {\bibinfo {volume} {40}},\ \bibinfo {pages} {3880}
  (\bibinfo {year} {1989})}\BibitemShut {NoStop}%
\bibitem [{\citenamefont {Belitz}\ \emph {et~al.}(2005)\citenamefont {Belitz},
  \citenamefont {Kirkpatrick},\ and\ \citenamefont {Votja}}]{Belitz05RMP}%
  \BibitemOpen
  \bibfield  {author} {\bibinfo {author} {\bibfnamefont {D.}~\bibnamefont
  {Belitz}}, \bibinfo {author} {\bibfnamefont {T.~R.}\ \bibnamefont
  {Kirkpatrick}}, \ and\ \bibinfo {author} {\bibfnamefont {T.}~\bibnamefont
  {Votja}},\ }\href@noop {} {\bibfield  {journal} {\bibinfo  {journal} {Rev.
  Mod. Phys.}\ }\textbf {\bibinfo {volume} {77}},\ \bibinfo {pages} {579}
  (\bibinfo {year} {2005})}\BibitemShut {NoStop}%
\bibitem [{\citenamefont {Law}\ \emph {et~al.}(1990)\citenamefont {Law},
  \citenamefont {Segre}, \citenamefont {Gammon},\ and\ \citenamefont
  {Sengers}}]{Law90PRA}%
  \BibitemOpen
  \bibfield  {author} {\bibinfo {author} {\bibfnamefont {B.~M.}\ \bibnamefont
  {Law}}, \bibinfo {author} {\bibfnamefont {P.~N.}\ \bibnamefont {Segre}},
  \bibinfo {author} {\bibfnamefont {R.~W.}\ \bibnamefont {Gammon}}, \ and\
  \bibinfo {author} {\bibfnamefont {J.~V.}\ \bibnamefont {Sengers}},\
  }\href@noop {} {\bibfield  {journal} {\bibinfo  {journal} {Phys. Rev. A}\
  }\textbf {\bibinfo {volume} {41}},\ \bibinfo {pages} {816} (\bibinfo {year}
  {1990})}\BibitemShut {NoStop}%
\bibitem [{\citenamefont {Segre}\ \emph {et~al.}(1992)\citenamefont {Segre},
  \citenamefont {Gammon}, \citenamefont {Sengers},\ and\ \citenamefont
  {Law}}]{Segre92PRA}%
  \BibitemOpen
  \bibfield  {author} {\bibinfo {author} {\bibfnamefont {P.~N.}\ \bibnamefont
  {Segre}}, \bibinfo {author} {\bibfnamefont {R.~W.}\ \bibnamefont {Gammon}},
  \bibinfo {author} {\bibfnamefont {J.~V.}\ \bibnamefont {Sengers}}, \ and\
  \bibinfo {author} {\bibfnamefont {B.~M.}\ \bibnamefont {Law}},\ }\href@noop
  {} {\bibfield  {journal} {\bibinfo  {journal} {Phys. Rev. A}\ }\textbf
  {\bibinfo {volume} {45}},\ \bibinfo {pages} {714} (\bibinfo {year}
  {1992})}\BibitemShut {NoStop}%
\bibitem [{\citenamefont {Segre}\ \emph {et~al.}(1993)\citenamefont {Segre},
  \citenamefont {Gammon},\ and\ \citenamefont {Sengers}}]{Segre93PRE}%
  \BibitemOpen
  \bibfield  {author} {\bibinfo {author} {\bibfnamefont {P.~N.}\ \bibnamefont
  {Segre}}, \bibinfo {author} {\bibfnamefont {R.~W.}\ \bibnamefont {Gammon}}, \
  and\ \bibinfo {author} {\bibfnamefont {J.~V.}\ \bibnamefont {Sengers}},\
  }\href@noop {} {\bibfield  {journal} {\bibinfo  {journal} {Phys. Rev. E}\
  }\textbf {\bibinfo {volume} {47}},\ \bibinfo {pages} {1026} (\bibinfo {year}
  {1993})}\BibitemShut {NoStop}%
\bibitem [{\citenamefont {Li}\ \emph {et~al.}(1994)\citenamefont {Li},
  \citenamefont {Segre}, \citenamefont {Gammon},\ and\ \citenamefont
  {Sengers}}]{Li94PA}%
  \BibitemOpen
  \bibfield  {author} {\bibinfo {author} {\bibfnamefont {W.~B.}\ \bibnamefont
  {Li}}, \bibinfo {author} {\bibfnamefont {P.~N.}\ \bibnamefont {Segre}},
  \bibinfo {author} {\bibfnamefont {R.~W.}\ \bibnamefont {Gammon}}, \ and\
  \bibinfo {author} {\bibfnamefont {J.~V.}\ \bibnamefont {Sengers}},\
  }\href@noop {} {\bibfield  {journal} {\bibinfo  {journal} {Physica A}\
  }\textbf {\bibinfo {volume} {204}},\ \bibinfo {pages} {399} (\bibinfo {year}
  {1994})}\BibitemShut {NoStop}%
\bibitem [{\citenamefont {Vailati}\ and\ \citenamefont
  {Giglio}(1996)}]{Vailati96PRL}%
  \BibitemOpen
  \bibfield  {author} {\bibinfo {author} {\bibfnamefont {A.}~\bibnamefont
  {Vailati}}\ and\ \bibinfo {author} {\bibfnamefont {M.}~\bibnamefont
  {Giglio}},\ }\href@noop {} {\bibfield  {journal} {\bibinfo  {journal} {{Phys.
  Rev. Lett.}}\ }\textbf {\bibinfo {volume} {77}},\ \bibinfo {pages} {1484}
  (\bibinfo {year} {1996})}\BibitemShut {NoStop}%
\bibitem [{\citenamefont {Vailati}\ and\ \citenamefont
  {Giglio}(1997)}]{Vailati97Nature}%
  \BibitemOpen
  \bibfield  {author} {\bibinfo {author} {\bibfnamefont {A.}~\bibnamefont
  {Vailati}}\ and\ \bibinfo {author} {\bibfnamefont {M.}~\bibnamefont
  {Giglio}},\ }\href@noop {} {\bibfield  {journal} {\bibinfo  {journal} {Nature
  (London)}\ }\textbf {\bibinfo {volume} {390}},\ \bibinfo {pages} {262}
  (\bibinfo {year} {1997})}\BibitemShut {NoStop}%
\bibitem [{\citenamefont {Takacs}\ \emph {et~al.}(2011)\citenamefont {Takacs},
  \citenamefont {Vailati}, \citenamefont {Cerbino}, \citenamefont {Mazzoni},
  \citenamefont {Giglio},\ and\ \citenamefont {Cannell}}]{Takacs11PRL}%
  \BibitemOpen
  \bibfield  {author} {\bibinfo {author} {\bibfnamefont {C.~J.}\ \bibnamefont
  {Takacs}}, \bibinfo {author} {\bibfnamefont {A.}~\bibnamefont {Vailati}},
  \bibinfo {author} {\bibfnamefont {R.}~\bibnamefont {Cerbino}}, \bibinfo
  {author} {\bibfnamefont {S.}~\bibnamefont {Mazzoni}}, \bibinfo {author}
  {\bibfnamefont {M.}~\bibnamefont {Giglio}}, \ and\ \bibinfo {author}
  {\bibfnamefont {D.~S.}\ \bibnamefont {Cannell}},\ }\href@noop {} {\bibfield
  {journal} {\bibinfo  {journal} {{Phys. Rev. Lett.}}\ }\textbf {\bibinfo
  {volume} {106}},\ \bibinfo {pages} {244502} (\bibinfo {year}
  {2011})}\BibitemShut {NoStop}%
\bibitem [{\citenamefont {Cerbino}\ \emph {et~al.}(2015)\citenamefont
  {Cerbino}, \citenamefont {Sun}, \citenamefont {Donev},\ and\ \citenamefont
  {Vailati}}]{Cerbino15SR}%
  \BibitemOpen
  \bibfield  {author} {\bibinfo {author} {\bibfnamefont {R.}~\bibnamefont
  {Cerbino}}, \bibinfo {author} {\bibfnamefont {Y.}~\bibnamefont {Sun}},
  \bibinfo {author} {\bibfnamefont {A.}~\bibnamefont {Donev}}, \ and\ \bibinfo
  {author} {\bibfnamefont {A.}~\bibnamefont {Vailati}},\ }\href@noop {}
  {\bibfield  {journal} {\bibinfo  {journal} {Sci. Rep.}\ }\textbf {\bibinfo
  {volume} {5}},\ \bibinfo {pages} {14486} (\bibinfo {year}
  {2015})}\BibitemShut {NoStop}%
\bibitem [{\citenamefont {Kirkpatrick}\ \emph
  {et~al.}(2016{\natexlab{b}})\citenamefont {Kirkpatrick}, \citenamefont
  {Ortiz~de Z\'arate},\ and\ \citenamefont {Sengers}}]{Kirkpatrick16PRELM}%
  \BibitemOpen
  \bibfield  {author} {\bibinfo {author} {\bibfnamefont {T.~R.}\ \bibnamefont
  {Kirkpatrick}}, \bibinfo {author} {\bibfnamefont {J.~M.}\ \bibnamefont
  {Ortiz~de Z\'arate}}, \ and\ \bibinfo {author} {\bibfnamefont {J.~V.}\
  \bibnamefont {Sengers}},\ }\href@noop {} {\bibfield  {journal} {\bibinfo
  {journal} {Phys. Rev. E}\ }\textbf {\bibinfo {volume} {93}},\ \bibinfo
  {pages} {032117} (\bibinfo {year} {2016}{\natexlab{b}})}\BibitemShut
  {NoStop}%
\bibitem [{\citenamefont {Edwards}\ and\ \citenamefont
  {Singh}(1979)}]{Edwards79}%
  \BibitemOpen
  \bibfield  {author} {\bibinfo {author} {\bibfnamefont {S.~F.}\ \bibnamefont
  {Edwards}}\ and\ \bibinfo {author} {\bibfnamefont {P.}~\bibnamefont
  {Singh}},\ }\href@noop {} {\bibfield  {journal} {\bibinfo  {journal} {J.
  Chem. Soc.{,} Faraday Trans. 2}\ }\textbf {\bibinfo {volume} {75}},\ \bibinfo
  {pages} {1001} (\bibinfo {year} {1979})}\BibitemShut {NoStop}%
\bibitem [{Lem( Eq2)}]{Lemmon06JCED}%
  \BibitemOpen
  \href@noop {} {\  (\bibinfo {year} {See~Supplemental~Material~at
  http://link.aps.org/ for available data for thermodynamic and transport
  properties for neopentane and toluene, as well as the details of the
  calculation of the parameter $A$ in Eq.(2).})}\BibitemShut {NoStop}%
\bibitem [{\citenamefont {Talbot}\ \emph {et~al.}(2017)\citenamefont {Talbot},
  \citenamefont {Kotar}, \citenamefont {Parolini}, \citenamefont {Di~Michele},\
  and\ \citenamefont {Cicuta}}]{Talbot17NC}%
  \BibitemOpen
  \bibfield  {author} {\bibinfo {author} {\bibfnamefont {E.~L.}\ \bibnamefont
  {Talbot}}, \bibinfo {author} {\bibfnamefont {J.}~\bibnamefont {Kotar}},
  \bibinfo {author} {\bibfnamefont {L.}~\bibnamefont {Parolini}}, \bibinfo
  {author} {\bibfnamefont {L.}~\bibnamefont {Di~Michele}}, \ and\ \bibinfo
  {author} {\bibfnamefont {P.}~\bibnamefont {Cicuta}},\ }\href@noop {}
  {\bibfield  {journal} {\bibinfo  {journal} {Nature Communications}\ }\textbf
  {\bibinfo {volume} {8}},\ \bibinfo {pages} {15351} (\bibinfo {year}
  {2017})}\BibitemShut {NoStop}%
\bibitem [{\citenamefont {Suzuki}\ \emph {et~al.}(2013)\citenamefont {Suzuki},
  \citenamefont {Takano},\ and\ \citenamefont {Matsushita}}]{Suzuki13JCP}%
  \BibitemOpen
  \bibfield  {author} {\bibinfo {author} {\bibfnamefont {J.}~\bibnamefont
  {Suzuki}}, \bibinfo {author} {\bibfnamefont {A.}~\bibnamefont {Takano}}, \
  and\ \bibinfo {author} {\bibfnamefont {Y.}~\bibnamefont {Matsushita}},\
  }\href@noop {} {\bibfield  {journal} {\bibinfo  {journal} {{J. Chem. Phys.}}\
  }\textbf {\bibinfo {volume} {139}},\ \bibinfo {pages} {184904} (\bibinfo
  {year} {2013})}\BibitemShut {NoStop}%
\end{thebibliography}

\begin{thebibliography}{4}%
\makeatletter
\providecommand \@ifxundefined [1]{%
 \@ifx{#1\undefined}
}%
\providecommand \@ifnum [1]{%
 \ifnum #1\expandafter \@firstoftwo
 \else \expandafter \@secondoftwo
 \fi
}%
\providecommand \@ifx [1]{%
 \ifx #1\expandafter \@firstoftwo
 \else \expandafter \@secondoftwo
 \fi
}%
\providecommand \natexlab [1]{#1}%
\providecommand \enquote  [1]{``#1''}%
\providecommand \bibnamefont  [1]{#1}%
\providecommand \bibfnamefont [1]{#1}%
\providecommand \citenamefont [1]{#1}%
\providecommand \href@noop [0]{\@secondoftwo}%
\providecommand \href [0]{\begingroup \@sanitize@url \@href}%
\providecommand \@href[1]{\@@startlink{#1}\@@href}%
\providecommand \@@href[1]{\endgroup#1\@@endlink}%
\providecommand \@sanitize@url [0]{\catcode `\\12\catcode `\$12\catcode
  `\&12\catcode `\#12\catcode `\^12\catcode `\_12\catcode `\%12\relax}%
\providecommand \@@startlink[1]{}%
\providecommand \@@endlink[0]{}%
\providecommand \url  [0]{\begingroup\@sanitize@url \@url }%
\providecommand \@url [1]{\endgroup\@href {#1}{\urlprefix }}%
\providecommand \urlprefix  [0]{URL }%
\providecommand \Eprint [0]{\href }%
\providecommand \doibase [0]{http://dx.doi.org/}%
\providecommand \selectlanguage [0]{\@gobble}%
\providecommand \bibinfo  [0]{\@secondoftwo}%
\providecommand \bibfield  [0]{\@secondoftwo}%
\providecommand \translation [1]{[#1]}%
\providecommand \BibitemOpen [0]{}%
\providecommand \bibitemStop [0]{}%
\providecommand \bibitemNoStop [0]{.\EOS\space}%
\providecommand \EOS [0]{\spacefactor3000\relax}%
\providecommand \BibitemShut  [1]{\csname bibitem#1\endcsname}%
\let\auto@bib@innerbib\@empty
%</preamble>
\bibitem [{\citenamefont {Lemmon}\ and\ \citenamefont
  {Span}(2006)}]{lemmon2006short}%
  \BibitemOpen
  \bibfield  {author} {\bibinfo {author} {\bibfnamefont {E.~W.}\ \bibnamefont
  {Lemmon}}\ and\ \bibinfo {author} {\bibfnamefont {R.}~\bibnamefont {Span}},\
  }\href@noop {} {\bibfield  {journal} {\bibinfo  {journal} {Journal of
  Chemical {\&} Engineering Data}\ }\textbf {\bibinfo {volume} {51}},\ \bibinfo
  {pages} {785} (\bibinfo {year} {2006})}\BibitemShut {NoStop}%
\bibitem [{NIS()}]{NIST}%
  \BibitemOpen
  \href@noop {} {\enquote {\bibinfo {title} {National institute of standards
  and technology},}\ }\bibinfo {howpublished}
  {http://wtt-pro.nist.gov/}\BibitemShut {NoStop}%
\bibitem [{\citenamefont {Callen}(1960)}]{CallenThermodynamics}%
  \BibitemOpen
  \bibfield  {author} {\bibinfo {author} {\bibfnamefont {H.}~\bibnamefont
  {Callen}},\ }\href@noop {} {\emph {\bibinfo {title} {Thermodynamics}}}\
  (\bibinfo  {publisher} {John Wiley {\&} Sons, INC},\ \bibinfo {year}
  {1960})\BibitemShut {NoStop}%
\bibitem [{web()}]{webbookNIST}%
  \BibitemOpen
  \href@noop {} {\enquote {\bibinfo {title} {National institute of standards
  and technology},}\ }\bibinfo {howpublished}
  {http://webbook.nist.gov/}\BibitemShut {NoStop}%
\end{thebibliography}
%\begin{thebibliography}{10}

%merlin.mbs apsrev4-1.bst 2010-07-25 4.21a (PWD, AO, DPC) hacked
%Control: key (0)
%Control: author (72) initials jnrlst
%Control: editor formatted (1) identically to author
%Control: production of article title (-1) disabled
%Control: page (0) single
%Control: year (1) truncated
%Control: production of eprint (0) enabled
%

\widetext
\clearpage
\begin{center}
\textbf{\large Supplemental Materials: \\ Giant Casimir Non-Equilibrium Forces Drive Coil-to-Globule Transition in Polymers}
\end{center}
%%%%%%%%%% Merge with supplemental materials %%%%%%%%%%
%%%%%%%%%% Prefix a "S" to all equations, figures, tables and reset the counter %%%%%%%%%%
\setcounter{equation}{0}
\setcounter{figure}{0}
\setcounter{table}{0}
\setcounter{page}{6}
\makeatletter
\renewcommand{\theequation}{S\arabic{equation}}
\renewcommand{\thefigure}{S\arabic{figure}}
%\renewcommand{\bibnumfmt}[1]{[S#1]}
%\renewcommand{\citenumfont}[1]{S#1}
%\title{Supplementary Information: \\ Giant Casimir Non-Equilibrium Forces Drive Coil-to-Globule Transition in Polymers}

% repeat the \author .. \affiliation  etc. as needed
% \email, \thanks, \homepage, \altaffiliation all apply to the current
% author. Explanatory text should go in the []'s, actual e-mail
% address or url should go in the {}'s for \email and \homepage.
% Please use the appropriate macro foreach each type of information

% \affiliation command applies to all authors since the last
% \affiliation command. The \affiliation command should follow the
% other information
% \affiliation can be followed by \email, \homepage, \thanks as well.
%Collaboration name if desired (requires use of superscriptaddress
%option in \documentclass). \noaffiliation is required (may also be
%used with the \author command).
%\collaboration can be followed by \email, \homepage, \thanks as well.
%\collaboration{}
%\noaffiliation

%\date{\today}

%\begin{abstract}
% insert abstract here
%\end{abstract}

% insert suggested PACS numbers in braces on next line
%\pacs{}
% insert suggested keywords - APS authors don't need to do this
%\keywords{}

%\maketitle must follow title, authors, abstract, \pacs, and \keywords
%\maketitle

% body of paper here - Use proper section commands
% References should be done using the \cite, \ref, and \label commands
\section{Sign of the Force and Value of the Critical Temperature Gradient in Neopentane and Toluene.}
In order to establish whether the long-range interactions induced by the thermal gradient are attractive or repulsive, we need to evaluate the sign of the parameter $A$ [see Eq.~(2) in the main text].
Because $C_P, \nu, D_T, (\gamma - 1) > 0$, the sign of $A$ depends on,
\begin{equation}
\bar{A} = \Big[ 1- \frac{1}{\alpha C_P} \Big( \frac{\partial C_P}{\partial T} \Big)_P + \frac{1}{\alpha^2} \Big( \frac{\partial \alpha}{\partial T} \Big)_P \Big].
\end{equation}
The value of $\bar{A}$ depends on the thermodynamic state of the system (namely, temperature and density) and on the nature of the solvent.
We consider the fluids neopentane and toluene.
The Helmholtz free energy [$F=F(V,T)$] for both of these solvents are provided in~\cite{lemmon2006short} as a function of volume and temperature.
Because to evaluate $\bar{A}$ we need derivatives of $C_P$ and $\alpha$ as a function of $T$ along an isobaric path, we derived an equation to extract $\Big(\frac{\partial C_P}{\partial T} \Big)_P$ and $\Big( \frac{\partial \alpha}{\partial T} \Big)_P$ as a function of terms involving only derivatives with respect to $T$ and $V$ along isochoric and isothermal paths, respectively.
Let $\phi=\phi(V,T)$ be any function of $V$ and $T$.
We show below that,
\begin{equation}
\Big(\frac{\partial \phi}{\partial T}\Big)_P = 
\Big(\frac{\partial \phi}{\partial T}\Big)_V -
\Big(\frac{\partial \phi}{\partial V}\Big)_T\frac{\Big[\frac{\partial}{\partial V} \Big(\frac{\partial F}{\partial T}\Big)_V \Big]_T}{\Big(\frac{\partial^2 F}{\partial V^2}\Big)_T}.
\label{Eq:FixPatVT}
\end{equation}
Plugging in Eq.~\ref{Eq:FixPatVT} $C_P$ and $\alpha$ in place for $\phi$, and using the data in~\cite{lemmon2006short} we obtain $\bar{A}$ (note that in~\cite{lemmon2006short} there is an expression for the molar heat capacity at constant pressure, but not for $\alpha$; we can use Eq.~\ref{Eq:FixPatVT} to extract $\alpha$ by substituting $\phi$ with $V$ and dividing the result by the volume; furthermore, to avoid confusion note that $\alpha$ is used in~\cite{lemmon2006short} to indicate the Helmholtz free energy).

\begin{figure}
\includegraphics{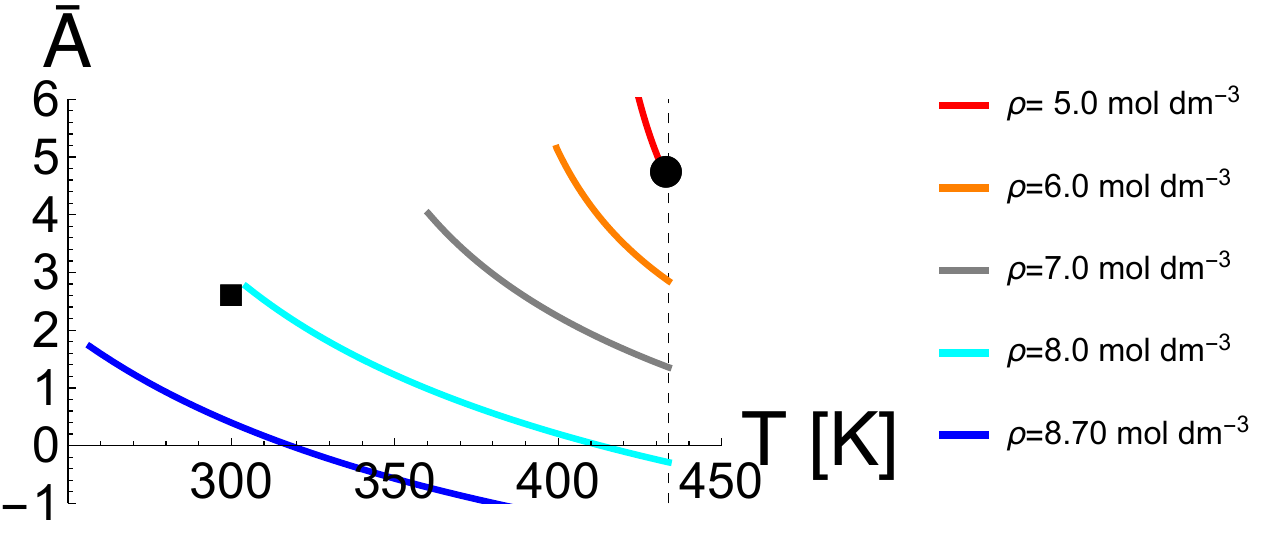}
\caption{(a) The value of $\bar{A}$ as a function of density and temperature for fluid neopentane.
Different lines correspond to different densities.
Each curve is shown in the temperature range corresponding to the fluid phase~\cite{webbookNIST}.
The dot and the square correspond to the values at which the evaluation of $A$ was performed; the dot has been obtained using ($T=433\Kelvin$, $\rho=4.2647~\mol\cdot\perdmcube$), and the value of $A$ for the thermodynamic state ($T=300\Kelvin$, $\rho=8.0927~\mol\cdot\perdmcube$) is shown as a square. 
The dashed vertical line shows the critical temperature, $T_c = 433.74 \Kelvin$~\cite{webbookNIST}.
\label{Fig:BarA}}
\end{figure}

The value of $\bar{A}$ for neopentane is shown in Fig.~\ref{Fig:BarA} as a function of temperature at different concentrations.
Clearly, we can find thermodynamic states for fluid neopentane with $\bar{A}>0$, which implies that the force is attractive and thus that for a supercritical thermal gradient the fluctuations in the NESS should induce the collapse of the polymer.
Note also that at large concentrations of neopentane it is possible to manipulate the sign of $\bar{A}$ by changing the overall temperature of the system.

\begin{figure}
\includegraphics{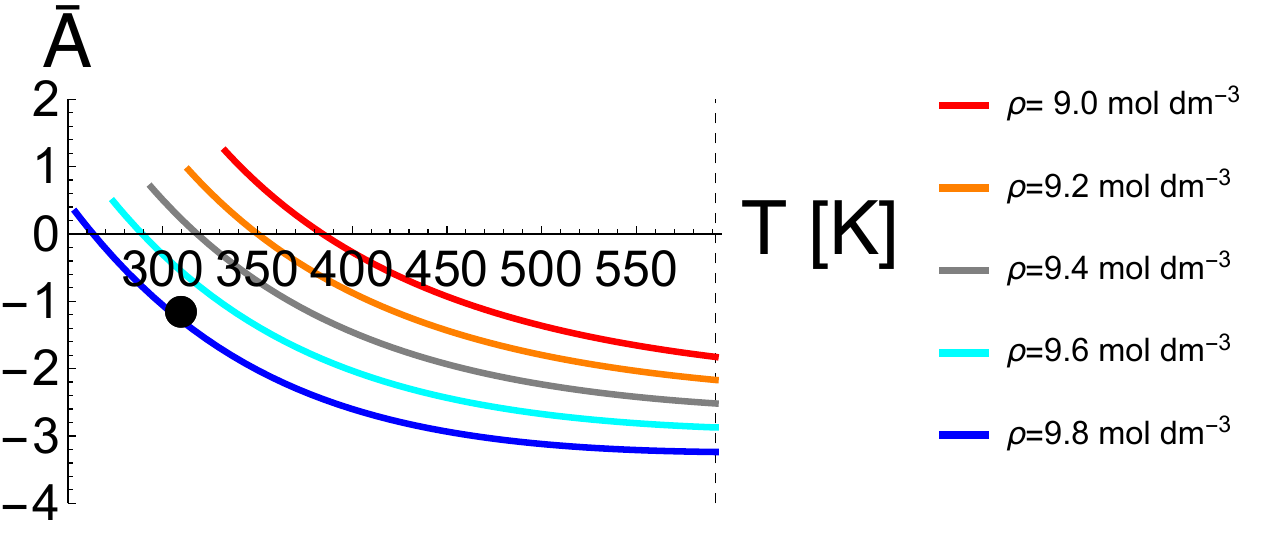}
\caption{Same as Fig.~\ref{Fig:BarA}, but for toluene.
The critical temperature is $T_c = 591.75~\Kelvin$~\cite{webbookNIST}). 
The dot shows the thermodynamic state ($T=310\Kelvin$, $\rho=9.75    ~\mol\cdot\perdmcube$).
\label{Fig:BarA_toluene}}
\end{figure}

In contrast, for toluene we can readily find $\bar{A}<0$, although small changes in the temperature change the sign of $\bar{A}$. 
Hence, small changes in the average temperature at constant thermal gradient can change the sign of the fluctuation-induced force from attractive to repulsive (see blue line in Fig.~\ref{Fig:BarA}, and Fig.~\ref{Fig:BarA_toluene}), thus effectively changing the quality of the solvent from bad to good, which is triggered by non-equilibrium fluctuations.

To provide a numerical estimate of the critical gradient, we need to compute the coefficient $A$, which requires estimating also the kinematic viscosity and thermal diffusivity.
We considered neopentane at room temperature ($T=300\Kelvin$) and $\rho=8.0927~\mol\cdot\perdmcube$ (see black square in Fig.~\ref{Fig:BarA}), and at a lower density ($\rho=4.2647~\mol\cdot\perdmcube$) close to the critical temperature ($T=433\Kelvin$) (see black circle in Fig.~\ref{Fig:BarA}).
The thermodynamic and transport coefficients were obtained from the online server of the National Institute of Standards and Technology (NIST)~\cite{NIST}.
The resulting parameters (see Table~\ref{Table:LiquidNeopentane}) yield $A\approx85~\pernmsquare$ in near-critical conditions, and $A\approx6.0\cdot10^{-2}~\pernmsquare$ at room temperature.
The larger value of $A$ close to the critical point accounts for the smaller critical gradient estimated in the main text (see Fig.~2).
Similarly, we report thermodynamics and transport coefficients for liquid toluene at $T=310\Kelvin$ and $\rho=9.75~\mol\cdot\perdmcube$ (see Table~\ref{Table:LiquidNeopentane}), which results in a negative value of $A$ ($A\approx-5.8\cdot10^{-3}~\pernmsquare$) indicating that in this thermodynamic state the presence of the thermal gradient does not modify the quality of the solvent.

\section{Derivation of Eq.~\ref{Eq:FixPatVT}.}
In order to derive Eq.~\ref{Eq:FixPatVT}, we consider the following three identities, which hold for any function $\phi = \phi(x,y)$, 
\begin{equation}
\Big(\frac{\partial \phi}{\partial u}\Big)_y = 
\Big(\frac{\partial \phi}{\partial x}\Big)_y\Big(\frac{\partial x}{\partial u}\Big)_y ; 
\label{Eq:A}
\end{equation}
\begin{equation}
\Big(\frac{\partial \phi}{\partial x}\Big)_v = 
\Big(\frac{\partial \phi}{\partial x}\Big)_y +
\Big(\frac{\partial \phi}{\partial y}\Big)_x\Big(\frac{\partial y}{\partial x}\Big)_v ; \\
\label{Eq:B}
\end{equation}
\begin{equation}
\Big(\frac{\partial \phi}{\partial x}\Big)_y = 
\frac{1}{\Big(\frac{\partial x}{\partial \phi}\Big)_y} .
\label{Eq:C}
\end{equation} 
%These three identities are readily derived considering using the chain rule with two new variables $u = u(x,y)$ and $v = v(x,y)$ 
%$\Big[$ that is $\Big(\frac{\partial \phi}{\partial u}\Big)_v = 
%\Big(\frac{\partial \phi}{\partial x}\Big)_y\Big(\frac{\partial x}{\partial u}\Big)_v +
%\Big(\frac{\partial \phi}{\partial y}\Big)_x\Big(\frac{\partial y}{\partial u}\Big)_v\Big]$
%, and then imposing $y=v$ (Eq.~\ref{Eq:A}), $x=u$ (Eq.~\ref{Eq:B}), or $v=y$ and $u=\phi$ (Eq.~\ref{Eq:C}).

Let $x=T$, $y=V$, and $v=P$, then from Eq.~\ref{Eq:B},
\begin{equation}
\Big(\frac{\partial \phi}{\partial T}\Big)_P = 
\Big(\frac{\partial \phi}{\partial T}\Big)_V +
\Big(\frac{\partial \phi}{\partial V}\Big)_T\Big(\frac{\partial V}{\partial T}\Big)_P .
\end{equation}
With the aid of the Maxwell relation $\Big(\frac{\partial V}{\partial T}\Big)_P = - \Big(\frac{\partial S}{\partial P}\Big)_T$ (see~\cite{CallenThermodynamics}) we get,
\begin{equation}
\Big(\frac{\partial \phi}{\partial T}\Big)_P = 
\Big(\frac{\partial \phi}{\partial T}\Big)_V -
\Big(\frac{\partial \phi}{\partial V}\Big)_T\Big(\frac{\partial S}{\partial P}\Big)_T,
\label{Eq:Inter}
\end{equation}
where $S$ is the entropy.

Let $\phi=S$, $x=V$, $u=P$, and $y=T$, then from Eq.~\ref{Eq:A},
\begin{equation}
\Big(\frac{\partial S}{\partial P}\Big)_T = 
\Big(\frac{\partial S}{\partial V}\Big)_T\Big(\frac{\partial V}{\partial P}\Big)_T,
\end{equation}
and using Eq.~\ref{Eq:C} for the second term on the r.h.s. we obtain,
\begin{equation}
\Big(\frac{\partial S}{\partial P}\Big)_T = 
\frac{\Big(\frac{\partial S}{\partial V}\Big)_T}{\Big(\frac{\partial P}{\partial V}\Big)_T}.
\label{Eq:S0}
\end{equation}
Let $F=F(V,T)$ be the Helmholtz free energy, then by plugging Eq.~\ref{Eq:S0} in Eq.~\ref{Eq:Inter} and substituting $P=  - \Big(\frac{\partial F}{\partial V}\Big)_T$ and $S = - \Big(\frac{\partial F}{\partial T}\Big)_V$ we obtain Eq.~\ref{Eq:FixPatVT}.

% Put \label in argument of \section for cross-referencing
%\section{\label{}}
\subsection{}
\subsubsection{}

% If in two-column mode, this environment will change to single-column
% format so that long equations can be displayed. Use
% sparingly.
%\begin{widetext}
% put long equation here
%\end{widetext}

% figures should be put into the text as floats.
% Use the graphics or graphicx packages (distributed with LaTeX2e)
% and the \includegraphics macro defined in those packages.
% See the LaTeX Graphics Companion by Michel Goosens, Sebastian Rahtz,
% and Frank Mittelbach for instance.
%
% Here is an example of the general form of a figure:
% Fill in the caption in the braces of the \caption{} command. Put the label
% that you will use with \ref{} command in the braces of the \label{} command.
% Use the figure* environment if the figure should span across the
% entire page. There is no need to do explicit centering.

\clearpage

% \begin{figure}
% \includegraphics{}%
% \caption{\label{}}
% \end{figure}

% Surround figure environment with turnpage environment for landscape
% figure
% \begin{turnpage}
% \begin{figure}
% \includegraphics{}%
% \caption{\label{}}
% \end{figure}
% \end{turnpage}

\begin{table}
\caption{Parameters for liquid neopentane and toluene.
The first column indicates the thermodynamic or transport coefficient.
$\eta$ is the viscosity; the kinematic viscosity is $\nu = \eta/\rho_m$, with $\rho_m$ mass density.
$k$ is the thermal conductivity, which is related to the thermal diffusivity by $D_T = k/(\rho C_P)$.
$c_P$ and $c_V$ are the molar heat capacities at constant pressure and volume, respectively, and $C_P$ is the specific heat capacity.
The second column reports the values of the coefficients at $T=300\Kelvin$ and $\rho=8.0927~\mol\cdot\perdmcube$ for neopentane.
The coefficients for neopentane in near-critical conditions are reported in the third column ($T=433\Kelvin$, $\rho=4.2647~\mol\cdot\perdmcube$).
The last column shows the parameters for toluene at $T=310\Kelvin$ and $\rho=9.75    ~\mol\cdot\perdmcube$.
The symbol ``$^{\dagger}$" indicates a quantity obtained from the NIST website~\cite{NIST}.
If a coefficient was computed using the Helmholtz free energy from~\cite{lemmon2006short}, the symbol ``$^{\ddagger}$" is used.
If a coefficient has been obtained in two different ways, the result from NIST are displayed; the value yielded by the Helmholtz free energy from~\cite{lemmon2006short} is within the error bar reported.
The results from NIST have been obtained at fixed temperature and pressure, while those using the Helmholtz free energy imposing the density and temperature.
For both thermodynamic states (second and third column), the density computed with the NIST website was within error bar from the density used for the Helmholtz free energy.}
\begin{center}
\begin{tabular}{| l || c || c || c |}
\hline
                 &           Neopentane                                                          &    Neopentane                                                                 &   Toluene             \\
\hline
parameter & 
\begin{tabular}{c} $T=300\Kelvin$ \\ $\rho=8.0927~\mol\cdot\perdmcube$ \end{tabular} & 
\begin{tabular}{c} $T=433\Kelvin$ \\ $\rho=4.2647~\mol\cdot\perdmcube$ \end{tabular} & 
\begin{tabular}{c} $T=310\Kelvin$ \\ $\rho=9.75    ~\mol\cdot\perdmcube$ \end{tabular} \\
\hline 
\hline
$\eta^{\dagger}~~[\Pa\cdot\s]$                                                    & $(2.4141\pm0.0061)\cdot 10^{-4}$ & $(3.75\pm0.23)\cdot 10^{-5}$  & $(7.90\pm0.13)\cdot 10^{-4}$ \\ 
\hline
$k^{\dagger}~~[\Watt\cdot\perm\cdot\perKelvin]$                       & $(9.333\pm0.015)\cdot 10^{-2}$     & $(6.68\pm0.30)\cdot 10^{-2}$  & $0.14856\pm0.00067$\\
\hline
$c_P^{\dagger,\ddagger}~[\Joule\cdot\permol\cdot\perKelvin]$ & $167.0\pm 2.0$                               & $770\pm 320$                          & $156.7 \pm 4.1$\\
\hline
$c_V^{\ddagger}~~[\Joule\cdot\permol\cdot\perKelvin]$            & $124.7$                                           & $176$                                       & $122.2$ \\
\hline
$\alpha^{\dagger,\ddagger}~[\perKelvin]$                                  & $(1.898\pm0.051)\cdot 10^{-3}$      & $(7.6\pm4.4)\cdot 10^{-2}$       & $(7.97\pm0.17)\cdot 10^{-4}$\\
\hline
\end{tabular}
\end{center}
\label{Table:LiquidNeopentane}
\end{table}%

\clearpage

\end{document}